\documentclass[preprint,12pt]{elsarticle}

\usepackage{amssymb}
\usepackage{epstopdf}
\usepackage{graphicx}
\usepackage{caption}
\usepackage{subcaption}
\usepackage{booktabs}
\usepackage{url}

\journal{Astroparticle Physics}

\begin{document}

\begin{frontmatter}



\title{Constraint on the axion-electron coupling constant and the neutrino magnetic dipole moment using the tip-RGB 
luminosity of 50 globular clusters.}


\author{S. Arceo-D\'iaz$^{1}$, K. -P. Schr\"oder$^{2}$, K. Zuber$^{3}$, D. 
Jack$^{2}$ and E. E. Bricio Barrios$^{1}$}
\address{$^{1}$Instituto Tecnol\'ogico de Colima, C. P. 28976, Villa de Alvarez, Colima, M\'exico.}
\address{$^{2}$Departamento de Astronom\'ia, Universidad de 
Guanajuato, C. P. 144, 36000, Guanajuato, Gto, M\'exico.
\textbf{email:}[santiago.arceo@itcolima.edu.mx}
\address{$^{3}$Institut f\"ur Kern- und Teilchenphysik,
Zellescher Weg 19, 01069, Dresden, Germany.}

\begin{abstract}

The current constrains in the neutrino magnetic dipole moment and axion-electron coupling constant are tested by estimating the maximum increment in bolometric luminosity, induced by the anomalous energy sink during the tip-RGB, that can be achieved by stellar tracks without entering in contradiction with the current observational calibration for the tip-RGB from globular clusters covering $\mathrm{-1.95\leq[M/H]\leq+0.04}$. When each energy sink is introduced into stellar models separately, the evidence gathered from the 50 most populated clusters suggests  $\mathrm{\alpha_{ae}\leq0.5\times10^{-26}}$ and $\mathrm{\mu_{\nu}\leq2.2\times10^{-12}\mu_{B}}$, while when they are assumed to occur simultaneously, these constraints are lowered down by 50\% and 25\%, respectively. 

\end{abstract}


\end{frontmatter}



\section{Introduction}

Neutrinos are already a canonical ingredient in stellar evolution. These particles, mostly known for being produced by nuclear fusion during the main-sequence or by the URCA reactions, at the onset of supernova explosions, can also be emitted by KeV-energy (``thermal") processes during the red giant branch (RGB). Although neutrinos are light particles, their corresponding flux steals a important amount of energy from the stellar interior and this has a large impact on defining the mass of the degenerate stellar core and on the overall bolometric luminosity prior to the helium flash \cite{BPS_1962,I_1996}. Certain properties of neutrinos are still unknown: if they are Dirac or Majorana particles \cite{C_1999}, the existence of ''sterile neutrinos'' \cite{A_2016}, the character of the neutrino mass spectrum \cite{P_2005} or if these particles do indeed have a magnetic dipole moment.\\

The amount of energy loss during the red giant branch could be further enhanced if the existence of other weakly interacting particles is confirmed by experiments. Axions, originally proposed as a way to solve CP-symmetry breaking problem \cite{PQ_1977}, are one example. Through their coupling with electrons, axions could be produced by the Compton and Bremsstrahlung processes and, if their emission rate is large enough, the resulting flux of energy would affect physical conditions during several stellar phases by cooling down the stellar interior. Axions are still considered theoretical but recent works have tried to prove their existence by determining the magnitude to their coupling to photons and electrons \cite{GGD_2018, ODGM_2018}.\\

Astrophysics provides an indirect method to constrain non-standard energy losses, either caused by neutrinos or axions. The bolometric luminosity of red giant stars is mostly dominated by the mass of their cores. The heavier the core gets the brighter the star becomes, as the surrounding hydrogen burning shell is forced by hydrostatic equilibrium to produce more energy to compensate for its gravitational pull. This leads to a feedback situation in which hydrogen burning produces more helium, making the core more heavy, and this, in turn, leads to a brighter tip-RGB. The helium flash, the event that terminates the red giant phase, happens only until the density and temperature of the core reach the critical values necessary for starting helium fusion by the triple-alpha process \cite{KW_2012}. The energy loss delays the helium flash, implying that the core has more time to augment its mass and the star becomes progressively brighter \cite{R_1998}. Raffelt et al. \cite{R_1990, R_1990B, R_1995} concluded that an excess in core mass of about $\mathrm{0.045M_{\odot}}$ would already lead to a clear conflict between the bolometric luminosity of stellar models and the observational evidence at the time. Recent studies based on Raffelt's method have proposed constraints for the magnetic dipole moment of neutrinos and the axion-electron coupling constant, based on the study on tip-RGB of the globular clusters M5, $\omega$-Centauri and M3 \cite{VX_2013B, ASZJ_2015, ODGM_2018}.\\ 

The question remains on how do these constraints are affected by the reported wide range of metallicity of most globular clusters, from which the tip-RGB luminosity depends. Here, we expand a previous work \cite{ASZJ_2015B} by also considering axion emission and by doubling the sample of globular clusters to fifty, extracted from the largest homogeneous infrared database by \cite{VFO_2004, VFO_2004B, Sol_2005, VFO_2007, VFO_2010}, instead of comparing the observational evidence from a single cluster against the predictions by stellar models. After describing the effect on both types of non-standard energy loss on stellar models (next section), we use the average tip-RGB absolute bolometric magnitude from the sample along with the absolute deviation of the median, as a robust estimator of dispersion, to put the constraint to both parameters.


\section{Non-standard energy losses and their theoretical rates}

This work uses the stellar evolution code created by Eggleton \cite{E_1971} as a computational basis. Our present version follows Pols et al. \cite{PTE_1995} for the 
prescription of the equation of state, nuclear reaction rates by Caughlan \& 
Fowler \cite{CF_1985}, electron conductivity according to Itoh et al. \cite{I_1983} and opacity tables, adapted specifically for the Eggleton code, by Chen \& Tout \cite{CT_2006}, following 
OPAL 96 \cite{IR_1996} for $\mathrm{\log_{10}{T/K} > 3.95}$ and Alexander \& Ferguson \cite{AF_1994} in the opposite range.\\

The energy lost by neutrino production in thermal processes is based on the analytical fits published Itoh et al. \cite{I_1992}, for the photo-, Bremsstrahlung-neutrino and pair-annihilation processes (the last one being irrelevant in at the density and temperature existing in the core of red giants) and the formula by Haft et al. \cite{H_1994} for plasmon decay. The effect of non-standard neutrino emission by modifying the emissivity of neutrinos by plasmon decay (in units of $\mathrm{erg\cdot g^{-1}\cdot s^{-1}}$) accordingly to Raffelt et al \cite{R_1992}.\\

Axion production was introduced into stellar models by considering the formulas described by Raffelt et al. \cite{R_1995} for the Compton and Bremsstrahlung processes. Raffelt et. al \cite{R_1995} constrained the axion-electron coupling constant to $\mathrm{\alpha_{26}\sim 0.6}$ (in units of $10^{-26}$), inducing an overall increment of about $\mathrm{0.045M_{\odot}}$, enhancing the bolometric luminosity at the tip-RGB over the observational evidence at that time.\\

Mass loss was included in stellar models trough the reinterpretation of Reimers's original formula done by Schr\"oder \& Cuntz \cite{SC_2005}, as it correctly allows to reproduce the envelope mass of red giants in globular clusters with a single value for the mass-loss parameter $\mathrm{\eta=8\times 10^{-14}M_{\odot} \cdot yr^{-1}}$, unlike more modern prescriptions \cite{CS_2011}.

\subsection{Effect of enhanced neutrino and axion emission on the degenerate core}

Once we have included both non-standard processes, we analyzed their effect on the energy balance of stellar models in two different scenarios: i) enhanced neutrino emission and ii) normal neutrino emission combined with axion production and compared them against the canonical case. We constructed stellar tracks with $\mathrm{M_{i}=0.8-1.2M_{\odot}}$, $\mathrm{Z=0.0001-0.02}$ and hydrogen and helium mass fractions according to Pols et al. \cite{PTS_1997}. For each track, we defined the theoretical tip-RGB as the model in which helium luminosity reaches to about $\mathrm{10}$ $\mathrm{L_{\odot}}$. Between this point and the true helium flash, bolometric luminosity does not changes significantly, as long as the canonical and non-standard models have the same helium luminosity, its increment due to a non-zero magnetic dipole moment or axion emission is not affected by this choice \cite{ASZJ_2015}.\\

\begin{figure}
\footnotesize{\caption{Neutrino emissivity during the RGB of a stellar track with M\ensuremath{_{i}=1.0}M\ensuremath{_{\odot}} 
and \ensuremath{Z=0.001}. The canonical scenario (the set of solid lines) is compared against those in which the energy losses have been enhanced by a magnetic dipole moment (dashed lines) or axion emission (dot-dashed lines). The left panel shows the temporal evolution for the luminosity peak of each process (the age of the stellar track, from the beginning of the main-sequence to the tip-RGB, has been normalized). The right panel shows the radial variation of the emissivity in the stellar model at the tip-RGB.}}
 \vspace{5mm}
    \begin{subfigure}[b]{0.50\linewidth}
    \includegraphics[angle=-90, width=0.95\linewidth]{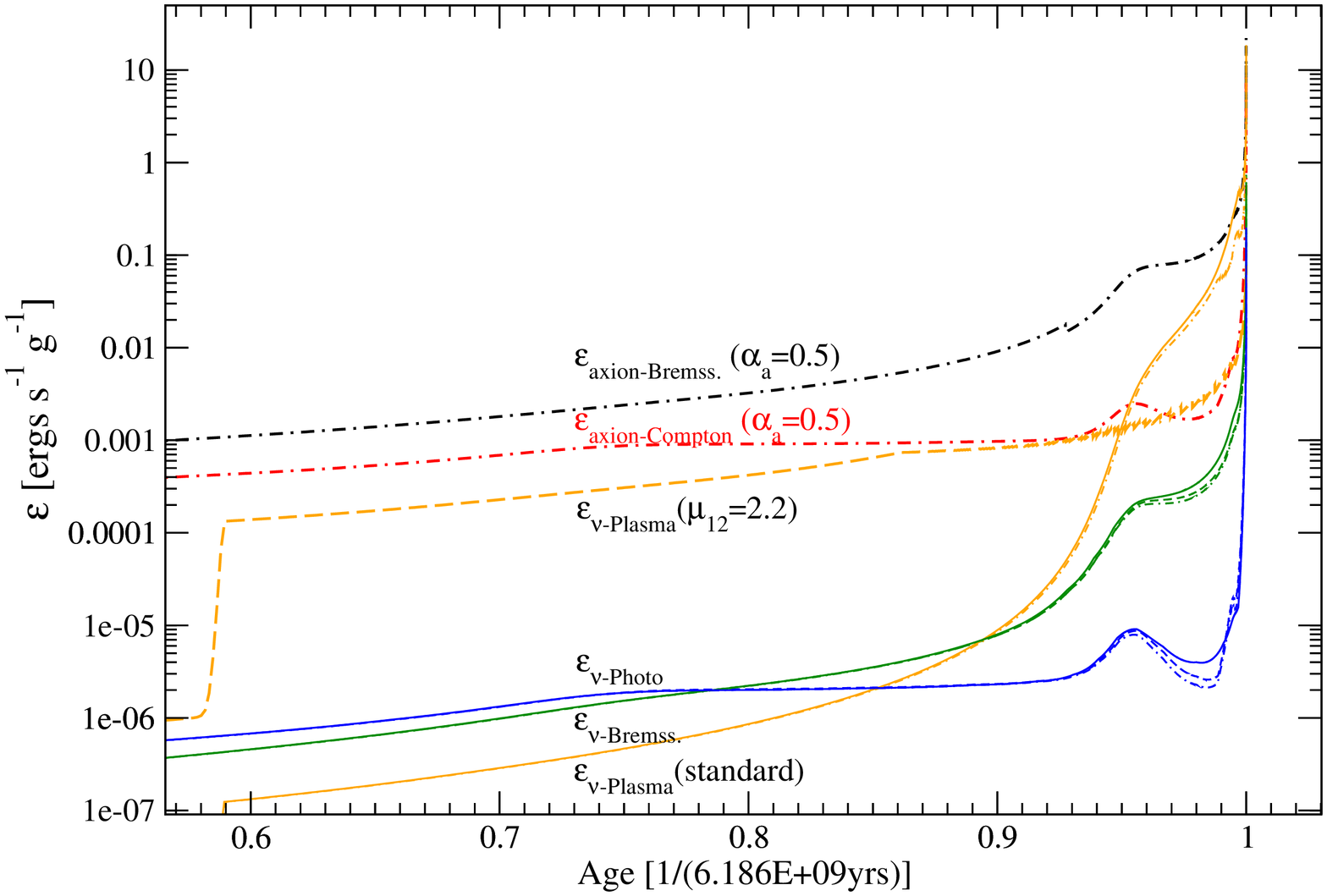}
  \end{subfigure}
  \quad
  \begin{subfigure}[b]{0.50\linewidth}
    \includegraphics[angle=-90, width=0.95\linewidth]{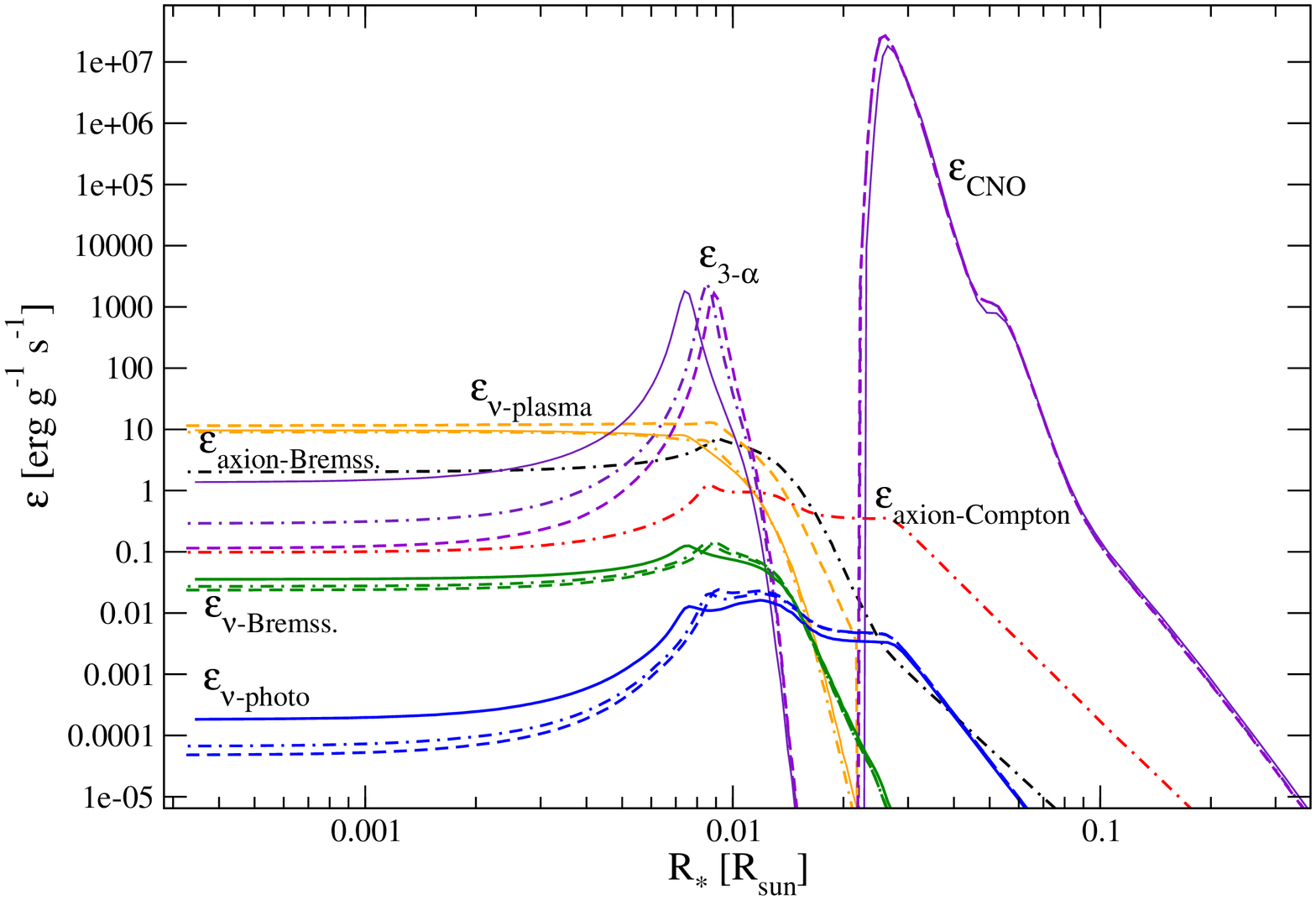}
  \end{subfigure}
  \label{fig1}
\end{figure}

The left panel on fig. 1 shows the temporal variation of the neutrino and axion emissivity during the last half of an stellar track defined by the initial values $\mathrm{M_{i}=1.0M_{\odot}}$ and $\mathrm{Z=0.001}$. On the canonical scenario, the emission of neutrinos by thermal processes steadily increases towards the tip-RGB and, at 0.94 of the total stellar age, there is an abrupt increment as the evolution speed is accelerated by the feedback between the core and the hydrogen burning shell. The majority of neutrinos are produced by the photo-neutrino process during the main-sequence until around 0.78 of the stellar age, when the early degeneracy of the stellar core makes easier to produce neutrinos by the Bremsstrahlung process. The increasing density of the stellar core accelerates neutrino production by plasmon decay and this becomes its largest source from 0.89 of the age to the onset of the helium flash.\\

In the scenario in which neutrino production is enhanced by a magnetic dipole moment (assumed here to be $\mu_{12}=2.2$) the main characteristics of the canonical scenario are mostly unchanged. The largest difference resides in the initial intensity of the process (dashed orange line on the left panel in fig. 1) as it starts being two orders of magnitude larger, steadily increasing towards the tip-RGB.\\ 

In the model including canonical neutrino production and axion emission (represented by the set of dot-dashed lines in fig. 1), the former is several orders of magnitude more intense since the beginning of the main-sequence, axion-Bremsstrahlung being the most important by an order of magnitude, and remain active until almost the very tip-RGB, when they become secondary to plasmon decay as an energy sink.\\ 

The variation of emissivity against stellar radius at the tip-RGB is shown in the right panel in fig. 1. Energy is produced by the onsetting 3-$\alpha$ fusion process on the stellar core and by the CNO-cycle burning hydrogen on the surrounding shell. Within the, almost, isothermal core, helium burning sets in at the point in which the shallow temperature gradient reaches its maximum (the main reason being that neutrino cooling, and degeneracy, gets increasingly stronger towards center). The degeneracy maintains neutrino production by plasmon decay almost constant inside the core (the maximum coinciding with the point in which helium starts to burn) and it decays rapidly at the exterior.\\ 

On the magnetic dipole moment scenario (dashed lines in fig. 1) the ignition point for helium gets displaced farther away from the stellar center, neutrino cooling is almost 30\% more intense than on the canonical scenario. On the transition region between the core and the hydrogen-burning shell, neutrino production due to plasmon decay does not fall abruptly and continues being important towards the base of the H-burning shell (implying that the decay on the emissivity of this process, due to the softening of degeneracy, is compensated by the magnetic dipole moment). These two features lead to the increment in bolometric luminosity shown in table 1.\\ 

In the axion scenario (dot-dashed lines in fig. 1) plasmon decay remains as the most efficient process for energy dissipation. Both axion processes dominate only after the point of helium ignition, axion-Bremsstrahlung dominates on the inner regions towards the stellar center (where density is high) and axion-Compton on the opposite direction, towards the H-burning shell, maintaining itself at a constant emissivity until the point in which hydrogen fusion reaches its maximum.


\subsection{Tip-RGB models with mass-loss and different neutrino dipole moment}

In this section we present our tip-RGB models and characterize their observable properties, using the tip-RGB models of stellar tracks with $\mathrm{M_{i}=1M_{\odot}}$ as 
representative cases, metallicity going from $\mathrm{Z=0.0001}$ to $\mathrm{Z=0.02}$ and the mass fractions of hydrogen and helium following \cite{PTS_1997} (see Table 1).  On each sub-table, the canonical tip-RGB is compared against the predictions considering either neutrino emission enhanced by the magnetic dipole moment (tables to the left) or by axion production (tables to the right). Each column represents the resulting tip-RGB when the current most restrictive constraints are taken to 25\%, 50\% or 100\% of their values).\\

\begin{table}[h!]
\scalebox{0.70}{
    \begin{subtable}[h]{0.8\textwidth}
\centering    
\begin{tabular}{lcccr}
	\toprule
$\mathrm{Z=0.0001}$ &  $\mu_{12}=0$    &   $\mu_{12}=0.55$  &
$\mu_{12}=1.1$  & $\mu_{12}=2.2$\\
\midrule
$\mathrm{\eta_{14}[M_{\odot} \cdot yr^{-1}]}$   & 8.00 & 7.70 & 6.50 & 5.50  \\
$\mathrm{M_{*} [M_{\odot}]}$ & 0.8804 & 0.8810 & 0.8912  &  0.8803 \\
$\mathrm{M_{c} [M_{\odot}]}$  & 0.4990  &  0.5010  & 0.5067 & 0.5237  \\
$\mathrm{\delta M_{c}[M_{\odot}]}$   & 0.0000  & 0.0022 & 0.0077 & 0.0247 \\
$\mathrm{L_{bol} [L_{\odot}]}$ & 1777 & 1830 & 1973 & 2418 \\
$\mathrm{T_{eff} [K]}$ & 4191  & 4184  & 4170    & 4112     \\
$\mathrm{R_{*} [R_{\odot}]}$ & 80 & 82 & 85 & 97 \\
\bottomrule
\end{tabular}
    \label{t4}
    \end{subtable}
~\quad
\begin{subtable}[h]{0.8\textwidth}
\centering    
\begin{tabular}{lcccr}
	\toprule
$Z=0.0001$ &  $\alpha_{26}=0$    &   $\alpha_{26}=0.125$  &
$\alpha_{26}=0.25$  & $\alpha_{26}=0.5$\\
\midrule
$\mathrm{\eta_{14}[M_{\odot} \cdot yr^{-1}]}$   & 8.00 & 7.70 & 6.50 & 5.60  \\
$\mathrm{M_{*} [M_{\odot}]}$ & 0.8804 & 0.8803 & 0.8814  &  0.8808 \\
$\mathrm{M_{c} [M_{\odot}]}$  & 0.4990  &  0.5060  & 0.5115 & 0.5220  \\
$\mathrm{\delta M_{c}[M_{\odot}]}$   & 0.0000  & 0.007 & 0.0125 & 0.023 \\
$\mathrm{L_{bol} [L_{\odot}]}$ & 1777 & 1944 & 2105  & 2377 \\
$\mathrm{T_{eff} [K]}$ & 4191  & 4168  & 4152 & 4125        \\
$\mathrm{R_{*} [R_{\odot}]}$ & 80 & 85 & 89 & 96 \\
\bottomrule
\end{tabular}
    \end{subtable}}
    \scalebox{0.70}{
        \begin{subtable}[h]{0.8\textwidth}
    \centering
    \begin{tabular}{lcccr}
	\toprule
$Z=0.001$  &  $\mu_{12}=0$    &   $\mu_{12}=0.55$  &
$\mu_{12}=1.1$  & $\mu_{12}=2.2$\\
\midrule
$\mathrm{\eta_{14} [M_{\odot} \cdot yr^{-1}]}$   & 8.00 & 7.70 & 6.80 & 5.30  \\
$\mathrm{M_{*} [M_{\odot}]}$ & 0.8433 & 0.8447 & 0.8425  &  0.8430 \\
$\mathrm{M_{c} [M_{\odot}]}$  & 0.4894  &  0.4912  & 0.4959 & 0.5104  \\
$\mathrm{\delta M_{c}[M_{\odot}]}$   & 0.0000  & 0.0018 & 0.0065 & 0.0207 \\
$\mathrm{L_{bol} [L_{\odot}]}$ & 2172 & 2220 & 2353 & 2784 \\
$\mathrm{T_{eff} [K]}$ & 3798  & 3791  & 3773    & 3711   \\
$\mathrm{R_{*} [R_{\odot}]}$ & 108 & 110 & 114 & 128 \\
\bottomrule
\end{tabular}
    \end{subtable}
~\quad 
    \begin{subtable}[h]{0.8\textwidth}
    \centering
    \begin{tabular}{lcccr}
	\toprule
$Z=0.001$ &  $\alpha_{26}=0$    &   $\alpha_{26}=0.125$  &
$\alpha_{26}=0.25$  & $\alpha_{26}=0.5$\\
\midrule
$\mathrm{\eta_{14} [M_{\odot} \cdot yr^{-1}]}$   & 8.00 & 7.20 & 6.50 & 5.60  \\
$\mathrm{M_{*} [M_{\odot}]}$ & 0.8433 & 0.8443 & 0.8469  &  0.8450 \\
$\mathrm{M_{c} [M_{\odot}]}$  & 0.4894  &  0.4952  & 0.5004 & 0.5092  \\
$\mathrm{\delta M_{c}[M_{\odot}]}$   & 0.0000  & 0.0058   & 0.011   & 0.0198  \\
$\mathrm{L_{bol} [L_{\odot}]}$ & 2172 & 2335 & 2485 & 2756  \\
$\mathrm{T_{eff} [K]}$ & 3798  & 3772  & 3751    & 3714   \\
$\mathrm{R_{*} [R_{\odot}]}$ & 108 & 114 & 118 & 127 \\
\bottomrule
\end{tabular}        
    \end{subtable}}
    \scalebox{0.70}{
        \begin{subtable}[h]{0.8\textwidth}
    \centering
     \begin{tabular}{lcccr}
	\toprule
$\mathrm{Z=0.01}$  &  $\mu_{12}=0$    &   $\mu_{12}=0.55$  &
$\mu_{12}=1.1$  & $\mu_{12}=2.2$\\
\midrule
$\mathrm{\eta_{14} [M_{\odot} \cdot yr^{-1}]}$   & 8.00 & 6.50 & 5.60 & 5.10 \\
$\mathrm{M_{*} [M_{\odot}]}$ & 0.7462 & 0.7464 & 0.7461  & 0.7480 \\
$\mathrm{M_{c} [M_{\odot}]}$  & 0.4794  & 0.4807   & 0.4843 & 0.4955  \\
$\mathrm{\delta M_{c} [M_{\odot}]}$   & 0.0000  & 0.0013 & 0.0049 & 0.0161 \\
$\mathrm{L_{bol} [L_{\odot}]}$ & 2572 & 2614 & 2727  & 3107 \\
$\mathrm{T_{eff} [K]}$   & 2970  & 2964   & 2934  & 2866  \\
$\mathrm{R_{*} [R_{\odot}]}$ & 192 & 194 & 203 & 226 \\
\bottomrule
\end{tabular}\label{t6}
\end{subtable}
~\quad
    \begin{subtable}[h]{0.8\textwidth}
    \centering
     \begin{tabular}{lcccr}
	\toprule
$Z=0.01$ &  $\alpha_{26}=0$    &   $\alpha_{26}=0.125$  &
$\alpha_{26}=0.25$  & $\alpha_{26}=0.5$\\
\midrule
$\mathrm{\eta_{14} [M_{\odot} \cdot yr^{-1}]}$   & 8.00 & 7.20 & 6.70 & 5.60 \\
$\mathrm{M_{*} [M_{\odot}]}$ & 0.7462 & 0.7475 & 0.7402  & 0.7471 \\
$\mathrm{M_{c} [M_{\odot}]}$  & 0.4794  & 0.4840   & 0.4882 & 0.4955  \\
$\mathrm{\delta M_{c} [M_{\odot}]}$   & 0.0000  & 0.0046 & 0.0080 & 0.0161 \\
$\mathrm{L_{bol} [L_{\odot}]}$ & 2572  & 2721  & 2858  & 3112 \\
$\mathrm{T_{eff} [K]}$   &  2970  & 2936   &  2901  & 2864  \\
$\mathrm{R_{*} [R_{\odot}]}$ & 192 & 203 & 212  & 227 \\
\bottomrule
\end{tabular}\label{t6}
    \end{subtable}}
    \scalebox{0.70}{\begin{subtable}[h]{0.8\textwidth}
    \centering
    \begin{tabular}{lcccr}
    \toprule
$\mathrm{Z=0.02}$  &  $\mu_{12}=0$    &   $\mu_{12}=0.55$  &
$\mu_{12}=1.1$  & $\mu_{12}=2.2$\\
\midrule
$\mathrm{\eta_{14} [M_{\odot} \cdot yr^{-1}]}$   & 8.00 & 7.75 & 7.20 & 5.55 \\
$\mathrm{M_{*} [M_{\odot}]}$ & 0.6979 & 0.7030 & 0.6983  & 0.6905 \\
$\mathrm{M_{c} [M_{\odot}]}$  & 0.4733  & 0.4744   & 0.4776  & 0.4876  \\
$\mathrm{\delta M_{c} [M_{\odot}]}$   & 0.0000  & 0.0011 & 0.0043 & 0.0143 \\
$\mathrm{L_{bol} [L_{\odot}]}$ & 2617 & 2653 & 2758 & 3108 \\
$\mathrm{T_{eff} [K]}$   &  2623  & 2611   & 2584  & 2536 \\
$\mathrm{R_{*} [R_{\odot}]}$ & 249 & 252 & 258 & 289 \\
\bottomrule
\end{tabular}
    \end{subtable}
~\quad    
    \begin{subtable}[h]{0.8\textwidth}
    \centering
    \begin{tabular}{lcccr}
    \toprule
$Z=0.02$ &  $\alpha_{26}=0$    &   $\alpha_{26}=0.125$  &
$\alpha_{26}=0.25$  & $\alpha_{26}=0.5$\\
\midrule
$\mathrm{\eta_{14} [M_{\odot} \cdot yr^{-1}]}$   & 8.00 & 7.60 & 6.40 & 5.40 \\
$\mathrm{M_{*} [M_{\odot}]}$ & 0.6979 & 0.6959 & 0.6971  & 0.7010 \\
$\mathrm{M_{c} [M_{\odot}]}$  & 0.4733  & 0.4770   & 0.4814  & 0.4881  \\
$\mathrm{\delta M_{c} [M_{\odot}]}$   & 0.0000  & 0.0033 & 0.0081 & 0.0150 \\
$\mathrm{L_{bol} [L_{\odot}]}$ & 2617 & 2759 & 2887 & 3133 \\
$\mathrm{T_{eff} [K]}$   &  2623  & 2578   & 2559  & 2535 \\
$\mathrm{R_{*} [R_{\odot}]}$ & 249 & 264 & 274 & 291 \\
\bottomrule
\end{tabular}
    \end{subtable}}\label{t3}
    \caption{Tip-RGB models with both mass-loss and non-standard neutrino 
emission.}
\end{table}

Although it would appear that the correct calibration of the mass loss rate during the RGB does not matter for the study of the effect of non-standard energy losses, via higher core masses, neutrino and axion cooling indirectly increase the total luminosity and influences the physical properties of the stellar envelope. In response to a heavier core, the envelope extends and leads to a lower effective temperature. This, in turn, would non-physically increase the mass-loss rate predicted by any given parametric prescription, even suppressing the helium flash \cite{ASZ_2013}. Due to this interaction, the mass-loss parameter had to be reduced from the calibration made by \cite{SC_2005} and the new values can be seen in the first line of each sub-table. The mass of the stellar models at the tip-RGB is shown in the second line and it does not varies more than 1\% between different stellar models having the same metallicity.\\

The third and fourth lines on table 1 show the mass of the degenerate 
helium core and its corresponding increment, driven either by a non-zero magnetic dipole moment or axion emission. The minimum increment over the canonical core mass, $\mathrm{\sim0.015M_{\odot}}$, proposed by Catelan et al. \cite{CDH_1996} to produce a observable difference in the bolometric magnitude of the tip-RGB is located between $\mu_{12}=1.1$-2.2 and $\alpha_{26}=0.25$-0.5 in all cases. There are slight variations due to initial composition and metallicity. We separate these from the intrinsic dependence on $\mu_{12}$ and $\alpha_{26}$ below.\\ 

The increment of the tip-RGB luminosity with non-standard energy losses results from a higher efficiency in the hydrogen-burning shell 
just above the degenerated helium core. In all our models, such a 
higher tip-RGB energy output leads to further envelope expansion, 
as is obvious from the respective stellar radii and temperatures (lines 6 and 7 on tables 1 and 2). In all our models with $\mu_{12}\geq 2$ and $\alpha_{26}\geq0.5$ the gains in stellar radii surpass $25\%$.\\

We derived an approximate parametric description for both, the mass of the 
helium core, $\mathrm{M_{c}}$, near the tip of the RGB and its increment by
non-standard neutrino cooling  and axion emission, based on our 
own models. To start with, the fit for the core mass of canonical models
(i.e., with standard neutrino cooling) $\mathrm{M_{c-std}}$, is: 

\begin{eqnarray}
\mathrm{M_{c-std}=0.4906-0.019M^{*}-0.008Z^{*}-0.22Y^{*}},
\end{eqnarray}

where $\mathrm{M^{*}}$, $\mathrm{Y^{*}}$ and $\mathrm{Z^{*}}$ are defined by:

\begin{eqnarray}
&\phantom{=}&\mathrm{M^{*}=M_{i}-0.95}\\
&\phantom{=}&\mathrm{Y^{*}=Y_{i}-0.242}\\
&\phantom{=}&\mathrm{Z^{*}=3 + \log_{10}{Z_{i}}},
\end{eqnarray} 

the sub-index $\mathrm{i}$ indicating initial values. In the scenario with a non-zero magnetic dipole moment, the non-standard increment in the core's mass is obtained by:

\begin{eqnarray}\label{coremass}
\mathrm{M_{c}=M_{c-std}+\delta M_{c}}. 
\end{eqnarray} We find that it depends mostly on $\mathrm{Z_{i}}$ and $\mathrm{M_{i}}$, whereas the initial value for the helium mass fraction ($\mathrm{Y_{i}}$) makes no difference. We characterize this dependence as:

\begin{eqnarray}
\mathrm{\delta M_{c}=\delta M_{\mu}^{*}(1-0.22Z^{*}+0.25M^{*})}, 
\end{eqnarray} where $\mathrm{\delta M_{\mu}^{*}}$ is the non-standard increment only due to a non-zero 
magnetic dipole moment: \begin{eqnarray}
\delta 
\mathrm{M_{\mu}^{*}=0.0267\left[(\mu_{1}^2+\mu_{12}^{2})^{0.5}-\mu_{1}-(\mu_{12}/\mu_{2}
)^ {1.5}\right]},
\end{eqnarray} with $\mu_{1}=1.2$ and $\mu_{2}=3.3$, as it was derived by 
Raffelt \cite{R_1992}. The bolometric luminosity can be calculated as a function of the mass of the core as:

\begin{equation}\label{LMc}
\mathrm{L_{bol}=1.58\times10^{5}M_{c}^{6}\times10^{0.77Y^{*}+0.12Z^{*}}}.
\end{equation} 

For axion emission the parametric increment in core mass is:
\begin{eqnarray}
\mathrm{\delta M_{c}=\delta M_{\alpha_{26}}^{*}(1-0.149Z^{*}-0.41M^{*})}, 
\end{eqnarray} where

\begin{eqnarray}
\mathrm{\delta M_{\alpha_{26}}^{*}=0.036\alpha_{26}+0.0015},
\end{eqnarray} while the mass-luminosity relation is given by:

\begin{eqnarray}\label{axionL}
\mathrm{L_{bol}=1.55\times 10^{5}M_{c}^{6}\times10^{0.77Y^{*}+0.12Z^{*}}}.
\end{eqnarray} These equations approximate the canonical and non-standard core mass and bolometric luminosity (as given by our models) for any stage on the ascend to the RGB in the plausible range of 
mass (0.8 to 1.2 $\mathrm{M_{\odot}}$), helium content and metallicity ($\mathrm{Z=0.0001}$ 
to $\mathrm{Z=0.02}$) with a maximum percentual error of about $1\%$.


\section{Comparison with observational data.}

\begin{figure}[h]
\footnotesize{\caption{The histogram of the sample (left upper panel) has a Gaussian profile centered around the mean $\langle$M$^{\mathrm{\scriptsize{tip}}}_{\mathrm{\scriptsize{obs}}}\rangle=-3.64$, while the average bolometric magnitude of non-standard models is 1.8-$\sigma$ away. A possible, systematic shift towards a brighter observational tip-RGB, caused by very bright clusters, is hinted after diving the sample in two groups (right upper panel). The mean bolometric magnitude and the median of 2000 resamplings by bootstrap are shown on the lower panels. Both suggest $-3.64$ as the bolometric magnitude of the tip-RGB, with an standard deviation of 0.04.}}\label{F03}
\vspace{6mm}
        \begin{subfigure}{0.46\linewidth}
                \includegraphics[width=0.9\linewidth]{samplehisto.eps}
        \end{subfigure}\hspace*{3mm}
         \begin{subfigure}{0.46\linewidth}
                \includegraphics[width=0.9\linewidth]{Histocompb.eps}   
        \end{subfigure}\vspace*{10mm}
        \begin{subfigure}{0.46\linewidth}
                \includegraphics[width=1.1\linewidth]{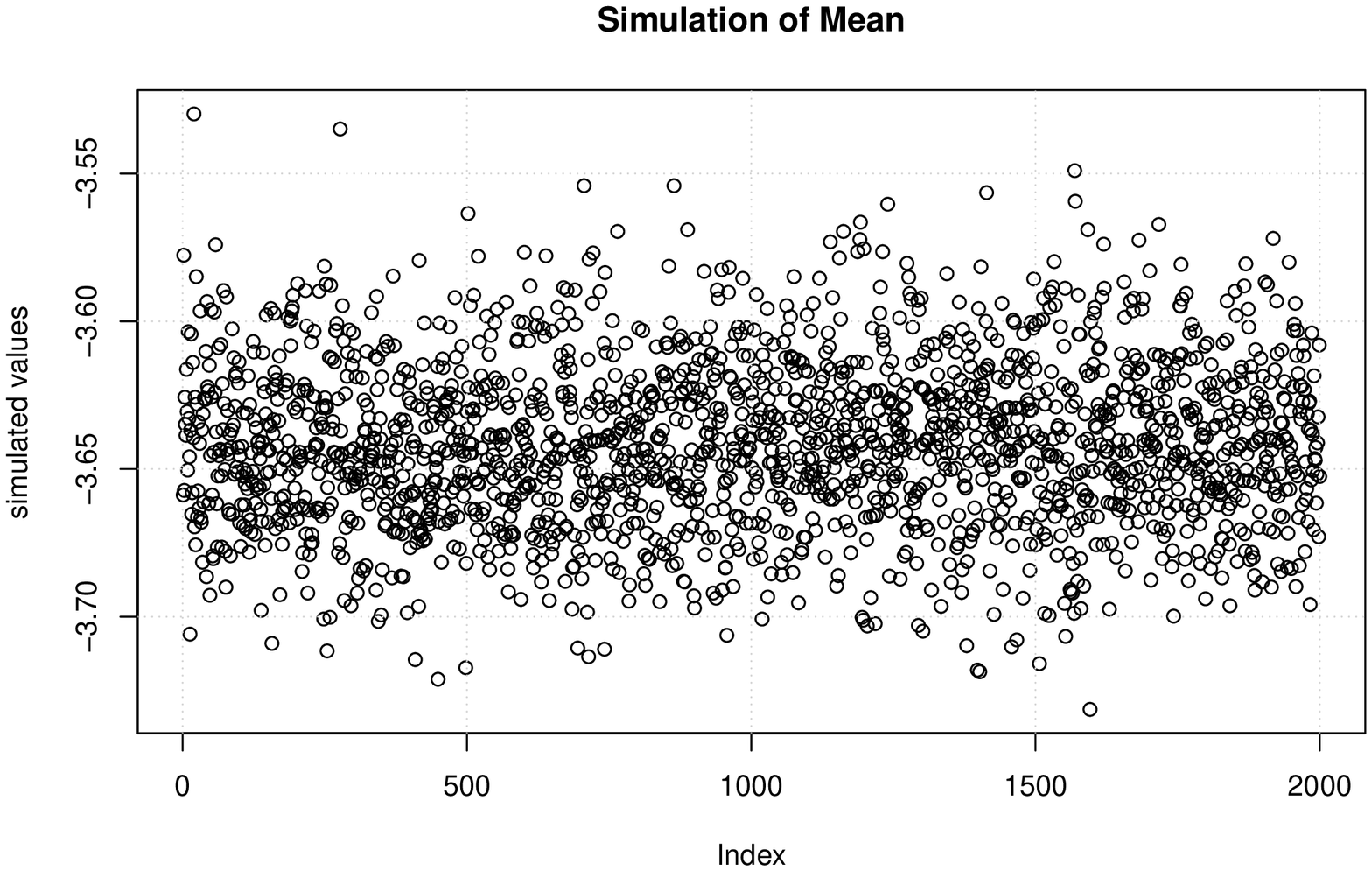}
        \end{subfigure}\hspace*{3mm}
        \begin{subfigure}{0.46\linewidth}
                \includegraphics[width=1.1\linewidth]{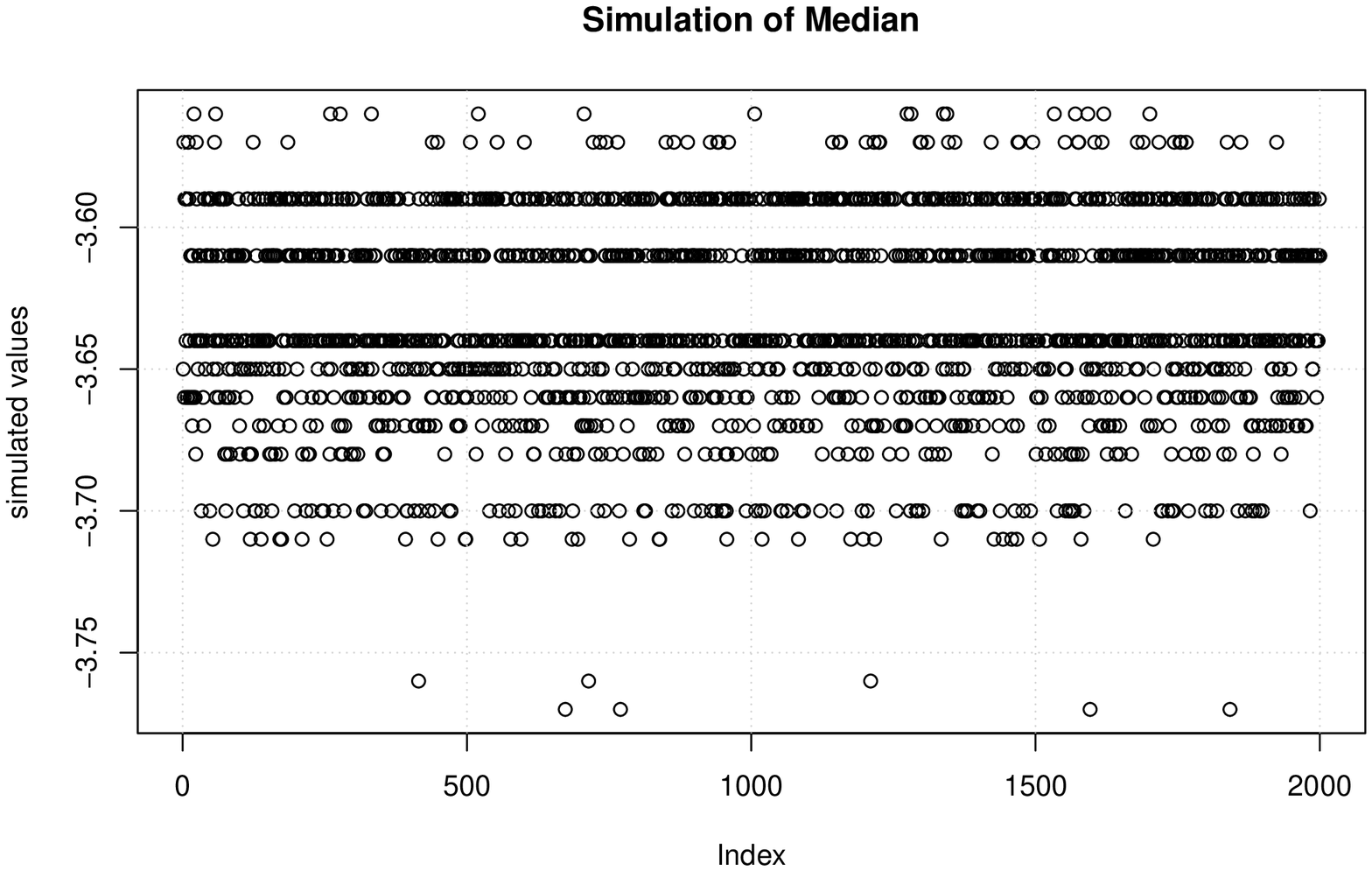}
        \end{subfigure}
         \end{figure}

Table 3 compares the calibrated and empirical tip-RGB bolometric magnitude of the globular clusters in our sample against the canonical and non-standard stellar models. The sample was extracted from the largest homogeneous NIR-database by selecting globular clusters with at least 30 stars on the two brightest bins below the RGB-tip, rendering the statistical uncertainty between this point and the brightest red giant to $\mathrm{\sigma_{s}\leq0.16}$ mag. Additionally, we selected only clusters in which the possible variations on the global metallicity would be small enough as to result on a single RGB (with the exception of $\omega$-Centauri, for which the multiple stellar populations are not relevant for the brightness of tip-RGB \cite{BFP_2001}). The empirical tip-bolometric magnitude (shown in column six) comes from the formula described by Valenti et al. \cite{VFO_2004}:
\begin{equation}
\mathrm{M_{emp}^{Tip}=-3.85+0.19[M/H]}.
\end{equation} All the stellar tracks have an initial mass M$_{i}=0.95$M$_{\odot}$ and Z matching the reported global metallicity [M/H] for each cluster. The particular choice of initial mass for stellar tracks corresponds to the central value within the range $\mathrm{M_{*}=0.9\pm0.15M_{\odot}}$ (covering the 0.06 mag. error bars shown in fig. 3 and more or less coinciding with the estimated upper and lower limits for the age of each cluster). The tip-RGB models correspond to the particular point in stellar tracks in which $\mathrm{L_{He}=10L_{\odot}}$.  For each cluster we used three stellar tracks: the canonical case and the two non-standard scenarios with $\mu_{12}=2.2$ and $\alpha_{26}=0.5$.\\
 
\begin{table}[p]
\centering
\scriptsize
\caption{Our sample. The bolometric magnitude for the tip-RGB of each cluster was extracted from: (a) Ferraro et al. \cite{F00}, (b to d) Valenti et al. \cite{VFO_2007, VFO_2004, VFO_2010} and (e) Sollima et al. \cite{Sol_2005}.}
\begin{tabular}{llllllll}
\toprule
\# & Cluster & [M/H] & $\mathrm{M_{obs}^{Tip}}$ & $\mathrm{M_{emp}^{Tip}}$ & $\mathrm{M_{0}^{Tip}}$ & $\mathrm{M_{\mu_{\nu}=2.2}^{tip}}$ & $\mathrm{M_{\alpha_{26}=0.5}^{tip}}$\\
\midrule
1 & M92$^{b}_{*}$          & -1.95 & -3.64$\pm$0.26 & -3.48 & -3.46 & -3.76 & -3.75 \\
2 & M15$^{a}_{*}$          & -1.91 & -3.55$\pm$0.20 & -3.49 & -3.47 & -3.77 & -3.75 \\
3 & M68$^{a}$              & -1.81 & -3.37$\pm$0.40 & -3.51 & -3.53 & -3.81 & -3.79 \\
4 & M30$^{a}$              & -1.71 & -3.70$\pm$0.35 & -3.52 & -3.52 & -3.82 & -3.81 \\
5 & M55$^{a}$              & -1.61 & -3.71$\pm$0.28 & -3.54 & -3.56 & -3.83 & -3.82 \\
6 & NGC6293$^{d}_{*}$      & -1.55 & -3.23$\pm$0.28 & -3.56 & -3.56 & -3.83 & -3.82 \\
7 & NGC6255$^{d}_{*}$      & -1.43 & -3.56$\pm$0.26 & -3.58 & -3.58 & -3.85 & -3.84 \\
8 & NGC6256$^{d}$          & -1.43 & -3.56$\pm$0.26 & -3.58 & -3.59 & -3.86 & -3.84 \\ 
9 & $\omega$-Cen.$^{e}_{*}$& -1.39 & -3.59$\pm$0.16 & -3.59 & -3.58 & -3.86 & -3.86 \\
10 & NGC6453$^{d}$         & -1.38 & -3.57$\pm$0.24 & -3.59 & -3.60 & -3.87 & -3.86 \\
11 & NGC6522$^{d}$         & -1.33 & -3.43$\pm$0.26 & -3.60 & -3.61 & -3.87 & -3.86 \\
12 & Djorg1$^{d}$          & -1.31 & -3.68$\pm$0.26 & -3.60 & -3.61 & -3.87 & -3.86 \\
13 & M10$^{b}$             & -1.25 & -3.61$\pm$0.26 & -3.61 & -3.65 & -3.89 & -3.88 \\
14 & NGC6273$^{d}_{*}$     & -1.21 & -3.56$\pm$0.26 & -3.62 & -3.61 & -3.87 & -3.87 \\
15 & NGC6401$^{d}_{*}$     & -1.20 & -3.42$\pm$0.26 & -3.62 & -3.63 & -3.88 & -3.87 \\
16 & M13$^{b}$             & -1.18 & -3.59$\pm$0.32 & -3.63 & -3.66 & -3.90 & -3.89 \\
17 & M3$^{b}_{*}$          & -1.16 & -3.61$\pm$0.24 & -3.63 & -3.62 & -3.88 & -3.88 \\
18 & NGC6540$^{d}$         & -1.10 & -3.56$\pm$0.26 & -3.64 & -3.63 & -3.88 & -3.87 \\
19 & Ter. 9$^{d}_{*}$      & -1.01 & -3.86$\pm$0.26 & -3.66 & -3.65 & -3.90 & -3.89 \\
20 & NGC6642$^{d}$         & -0.99 & -3.66$\pm$0.26 & -3.66 & -3.69 & -3.92 & -3.96 \\
21 & NGC6342$^{d}$         & -0.99 & -3.70$\pm$0.32 & -3.66 & -3.75 & -3.96 & -3.96 \\
22 & M4$^{a}$              & -0.94 & -3.67$\pm$0.22 & -3.67 & -3.70 & -3.92 & -3.92 \\
23 & HP1$^{d}$             & -0.91 & -3.56$\pm$0.26 & -3.68 & -3.69 & -3.92 & -3.91 \\
24 & M5$^{b}$              & -0.90 & -3.64$\pm$0.28 & -3.68 & -3.66 & -3.93 & -3.92 \\ 
25 & NGC6266$^{d}$         & -0.88 & -3.47$\pm$0.26 & -3.68 & -3.72 & -3.94 & -3.93 \\
26 & NGC288$^{c}$          & -0.85 & -3.80$\pm$0.25 & -3.69 & -3.71 & -3.92 & -3.95 \\
27 & NGC6265$^{d}$         & -0.80 & -3.56$\pm$0.26 & -3.70 & -3.68 & -3.94 & -3.93 \\
28 & NGC6638$^{d}_{*}$     & -0.78 & -3.88$\pm$0.35 & -3.70 & -3.68 & -3.93 & -3.92 \\
29 & M107$^{a}$            & -0.70 & -3.57$\pm$0.40 & -3.71 & -3.73 & -3.95 & -3.94 \\
30 & NGC6380$^{d}_{*}$     & -0.68 & -3.88$\pm$0.22 & -3.72 & -3.70 & -3.94 & -3.93 \\
31 & NGC6569$^{d}_{*}$     & -0.66 & -3.59$\pm$0.26 & -3.72 & -3.70 & -3.95 & -3.93 \\
32 & Ter. 3$^{d}_{*}$      & -0.63 & -3.47$\pm$0.26 & -3.73 & -3.74 & -3.96 & -3.95 \\
33 & NGC6539$^{d}$         & -0.60 & -3.77$\pm$0.26 & -3.74 & -3.74 & -3.96 & -3.95 \\
34 & 47-Tuc$^{a}_{*}$      & -0.59 & -3.71$\pm$0.19 & -3.74 & -3.70 & -3.96 & -3.95 \\
35 & NGC6637$^{d}_{*}$     & -0.57 & -3.34$\pm$0.31 & -3.74 & -3.71 & -3.95 & -3.94 \\
36 & NGC6304$^{d}_{*}$     & -0.56 & -3.59$\pm$0.33 & -3.74 & -3.71 & -3.95 & -3.94 \\ 
37 & M69$^{a}$             & -0.55 & -3.51$\pm$0.25 & -3.75 & -3.75 & -3.96 & -3.96 \\
38 & Ter. 2$^{d}_{*}$      & -0.53 & -3.81$\pm$0.26 & -3.75 & -3.74 & -3.96 & -3.95 \\
39 & NGC6752$^{c}$         & -0.53 & -3.65$\pm$0.28 & -3.75 & -3.66 & -3.89 & -3.88 \\  
40 & NGC6441$^{d}_{*}$     & -0.52 & -3.90$\pm$0.20 & -3.75 & -3.72 & -3.94 & -3.95 \\
41 & NGC6624$^{d}$         & -0.48 & -3.85$\pm$0.31 & -3.76 & -3.75 & -3.97 & -3.97 \\
42 & Djorg2$^{d}_{*}$      & -0.45 & -3.50$\pm$0.26 & -3.76 & -3.76 & -3.96 & -3.97 \\
43 & Ter. 6$^{d}_{*}$      & -0.43 & -3.89$\pm$0.26 & -3.77 & -3.75 & -3.96 & -3.96 \\
44 & NGC6388$^{d}_{*}$     & -0.42 & -3.76$\pm$0.26 & -3.77 & -3.72 & -3.96 & -3.96 \\
45 & NGC6440$^{d}_{*}$     & -0.40 & -3.82$\pm$0.21 & -3.77 & -3.73 & -3.96 & -3.96 \\
46 & NGC6316$^{d}$         & -0.38 & -3.77$\pm$0.25 & -3.78 & -3.77 & -3.98 & -3.98 \\
47 & NGC6553$^{d}$         & -0.36 & -3.86$\pm$0.27 & -3.78 & -3.72 & -3.92 & -3.93 \\
48 & Ter. 5$^{d}_{*}$      & -0.14 & -3.96$\pm$0.26 & -3.82 & -3.62 & -3.97 & -3.97 \\
49 & Lillier1$^{d}$        & -0.14 & -3.81$\pm$0.26 & -3.82 & -3.77 & -3.92 & -3.92 \\
50 & NGC6528$^{d}_{*}$     & +0.04 & -4.06$\pm$0.25 & -3.86 & -3.78 & -3.96 & -3.99 \\ 
\bottomrule
\end{tabular}
\label{sample}
\end{table} 

First we analyzed the sample, focusing on its similarity to a Gaussian distribution. According to the Anderson-Darling test, it is reasonable to assume that the data sample has a Gaussian profile if the statistical parameter A fulfills the condition $\mathrm{A^{2}\leq0.752}$ \cite{AD_1952}. This parameter is related to the sample size as: 
\begin{equation}
\mathrm{A^{2}=-N-S},
\end{equation} where S is given by
\begin{equation}
\mathrm{S=\sum_{i=1}^{N}\frac{2i-1}{N}\left[lnF(Y_{i})+ln(1-F(Y_{N+1-i}))\right]} 
\end{equation} and F is the cumulative Gaussian distribution function. Our sample gives $\mathrm{A^{2}=0.55}$.\\ 

The mean value for the absolute bolometric magnitude of the tip-RGB from the sample is $\langle$M$^{\mathrm{\scriptsize{tip}}}_{\mathrm{\scriptsize{obs}}}\rangle=-3.64$ (with an standard deviation $\sigma=0.17$). The histogram for the sample is shown in the left panel inside fig. 2. The majority of globular clusters (18) indicate an absolute bolometric magnitude for the tip-RGB around $\mathrm{M_{bol}^{tip}=-3.64}$ while 88\% of them (44) have $\mathrm{M_{bol}^{tip}}$ within the closed interval $[-3.90,-3.40]$. From the remaining six, only two have $\mathrm{M_{bol}^{tip}}<-3.90$, which would represent 5\% of the sample.\\

As more robust test on the spread of the tip-RGB in the sample, we followed Leys et al. \cite{L_2013} by using the median absolute deviation:

\begin{equation}
\mathrm{MAD=median_{\scriptsize{i}}(|M_{\scriptsize{i}}-\overline{M}|)},
\end{equation} were M$_{i}$ refers to the tip-RGB absolute bolometric magnitude of any cluster in the sample and $\mathrm{\bar{x}}$ is the median ($\mathrm{\overline{M}=-3.64}$). The median absolute deviation is a robust measure of central dispersion and is related to the population standard deviation by $\mathrm{\sigma_{pop}=1.482}$ MAD. With the data in our sample, the standard deviation of the population can be approximated to $\sigma_{\mathrm{\scriptsize{pop}}}=0.15$ mag. Similar values were obtained by the bootstrap technique \cite{E_1971}. After 2000 resamplings (shown in the lower panels in fig. 2) the estimated values for the population's mean and median are $\langle$M$_{\mathrm{\scriptsize{bol}}}^{\mathrm{\scriptsize{tip}}}\rangle$=3.64=$\overline{\mathrm{M}}_{\mathrm{\scriptsize{bol}}}^{\mathrm{\scriptsize{tip}}}$, both with standard deviations of 0.04. These tests allow to estimate $\langle$M$_{bol}\rangle=3.64\pm0.30$ and $\overline{\mathrm{M}}-3.64\pm0.30$ to a two-$\sigma$ confidence level, placing the mean absolute bolometric magnitude of our non-standard models, $\langle$M$^{\mathrm{tip}}_{\mu_{12}=2.2}\rangle=-3.92$ and $\langle$M$^{\mathrm{tip}}_{\alpha_{26}=0.5}\rangle=-3.91$, at 1.8-$\sigma$ away from the observational calibration (implying that the probability of finding any globular cluster with a brighter tip-RGB is only 5\%).\\

The biasing of metallicity in the distribution (if globular clusters with high [M/H] could have tip-RGB bolometric luminosity similar to that of non-standard models) was analyzed by dividing our sample in two groups gathered around $\omega$-Centauri and 47-Tucanae (set by Ferraro et al. \cite{F00} and Bellazzini et al. \cite{BFP_2001, BFS_2004} as pillars for the calibration of the tip-RGB). The histograms of both sub-samples is shown in the right panel in fig. 2. The profile of the first group is almost Gaussian, with eight clusters located around the mean and median at $-3.55$ and $-3.58$. The second group has two peaks: the highest between M$_{\mathrm{\scriptsize{bol}}}^{\mathrm{\scriptsize{tip}}}=-3.90$ and -3.80 (nine clusters) while the secondary peak (six clusters) coincides with the one in the first group. The bi-modal profile of 47-Tucanae's group can be advocated to systematic effects due to the high contamination on the bulge. Despite the clusters in the second group indeed have brighter tip-RGB, only tow (Terzan 5 and NGC 6528 achieve a similar bolometric magnitude as that of non-standard models). These globular clusters with an atypically bright tip-RGB, are in any case shifting the overall observational values to higher bolometric magnitudes. Any future improvement on the calibration for these would allow even tighter constraints on $\mu_{\nu}$.\\

\begin{figure}[p]
\footnotesize{\caption{Comparison between the observational tip-RGB (black 
asterisk), its empirical 
estimation (green square) and our canonical (red circle) and non-standard 
models, with 
\ensuremath{\mu_{12}=2.2} (upward-blue triangle) and $\alpha_{26}$ (purple diamonds). In the two upper panels, twelve clusters (inside blue boxes) can 
be used to constrain \ensuremath{\mu_{12}\leq2.2}. The lower panels show results with shifts of $0.1$ and $0.16$ mag}}\label{figure00}
\vspace{6mm}
        \begin{subfigure}{0.46\linewidth}
                \includegraphics[width=.8\linewidth, angle=-90]{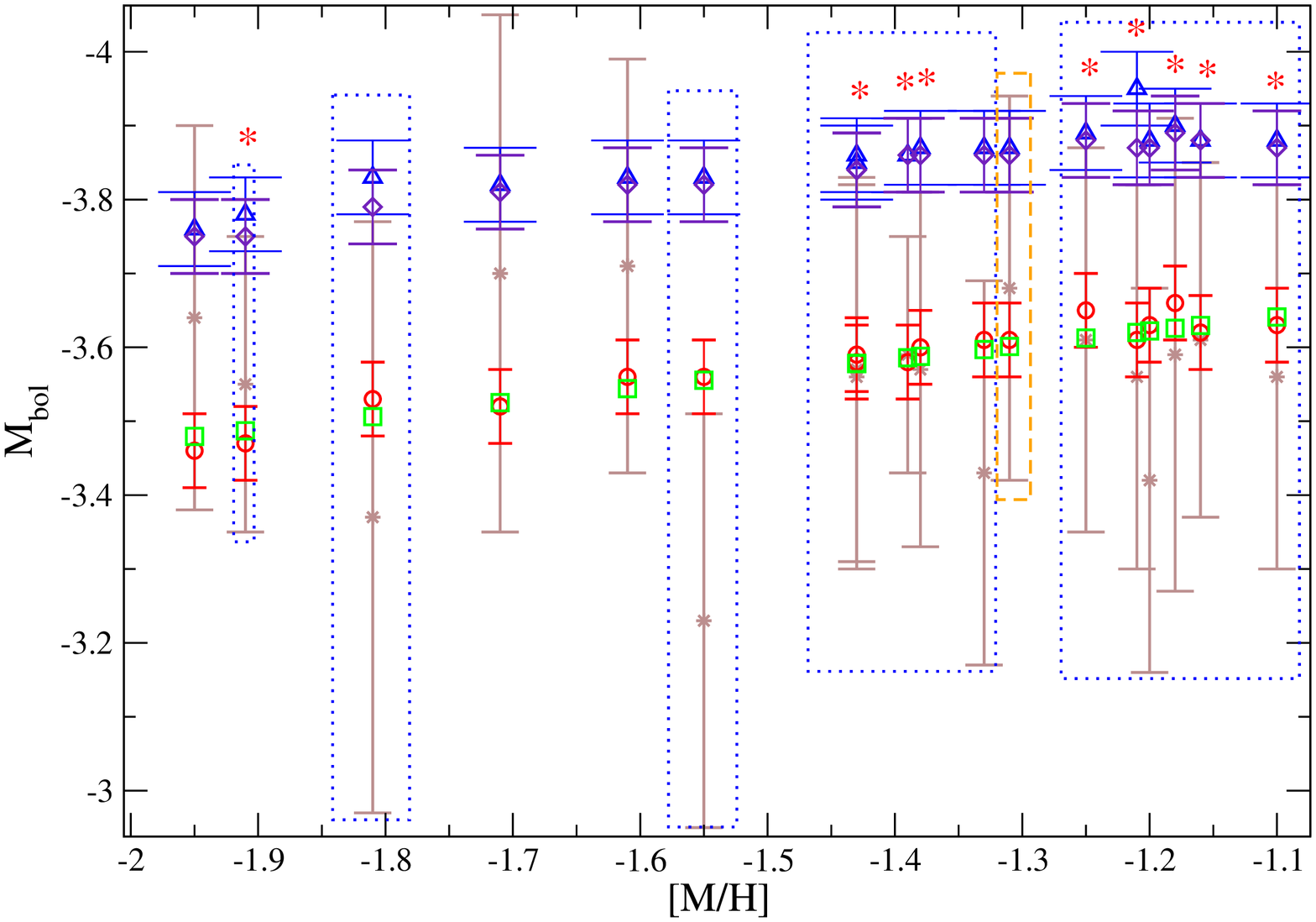}
        \end{subfigure}\hspace*{12mm}
         \begin{subfigure}{0.46\linewidth}
                \includegraphics[width=.8\linewidth, angle=-90]{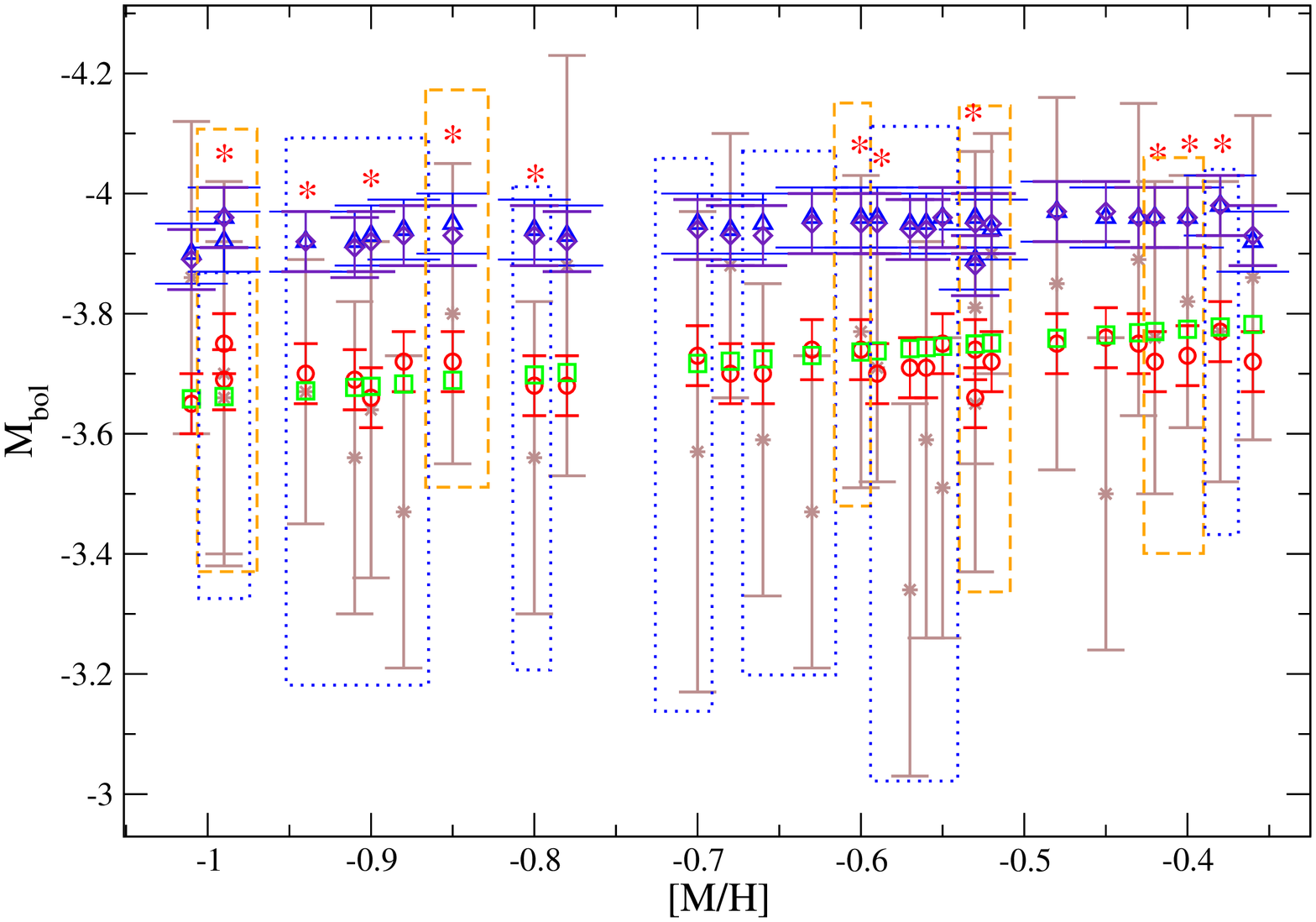}   
        \end{subfigure}\vspace*{10mm}
        \begin{subfigure}{0.46\linewidth}
                \includegraphics[width=.8\linewidth, angle=-90]{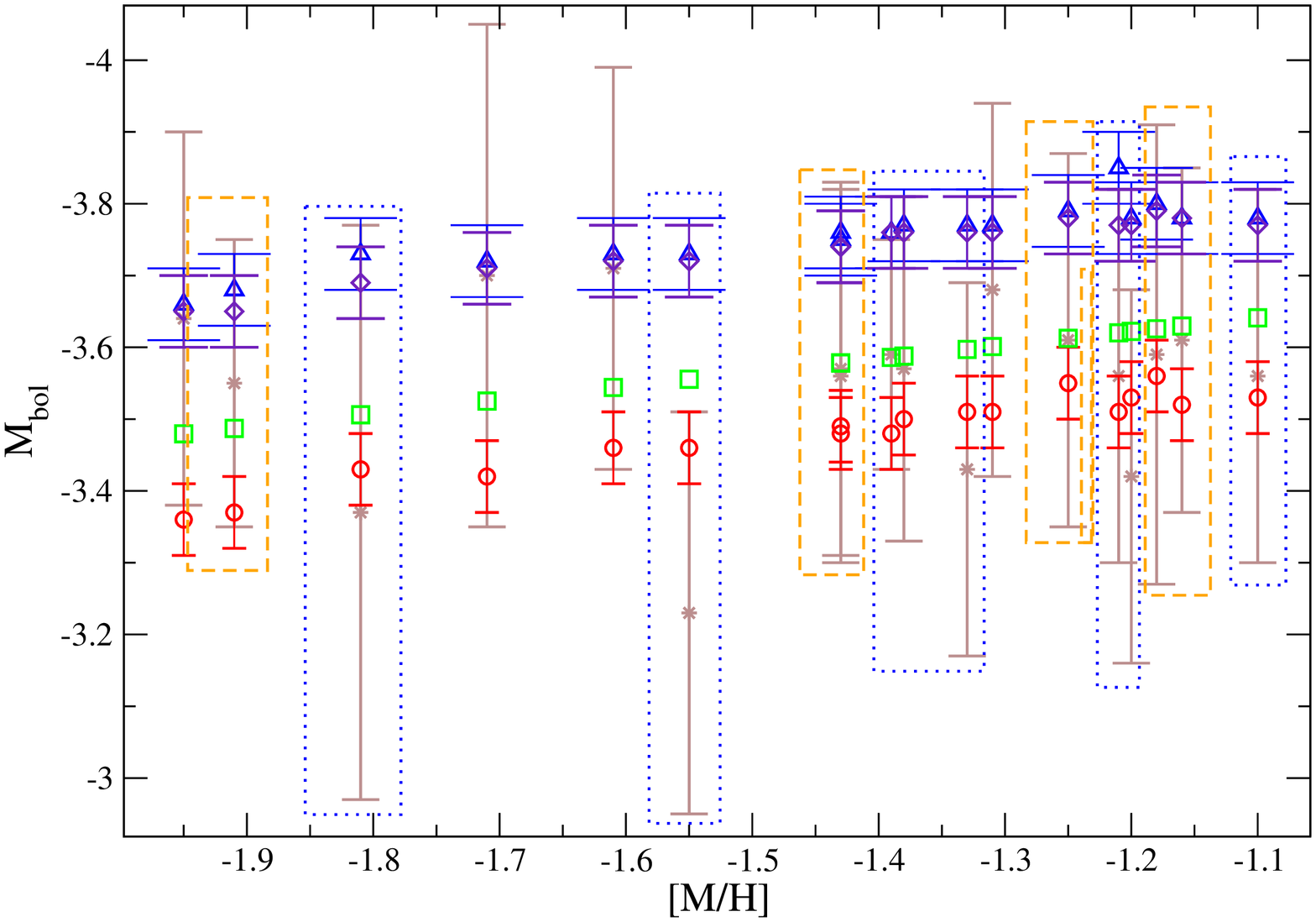}
        \end{subfigure}\hspace*{12mm}
        \begin{subfigure}{0.46\linewidth}
                \includegraphics[width=.8\linewidth, angle=-90]{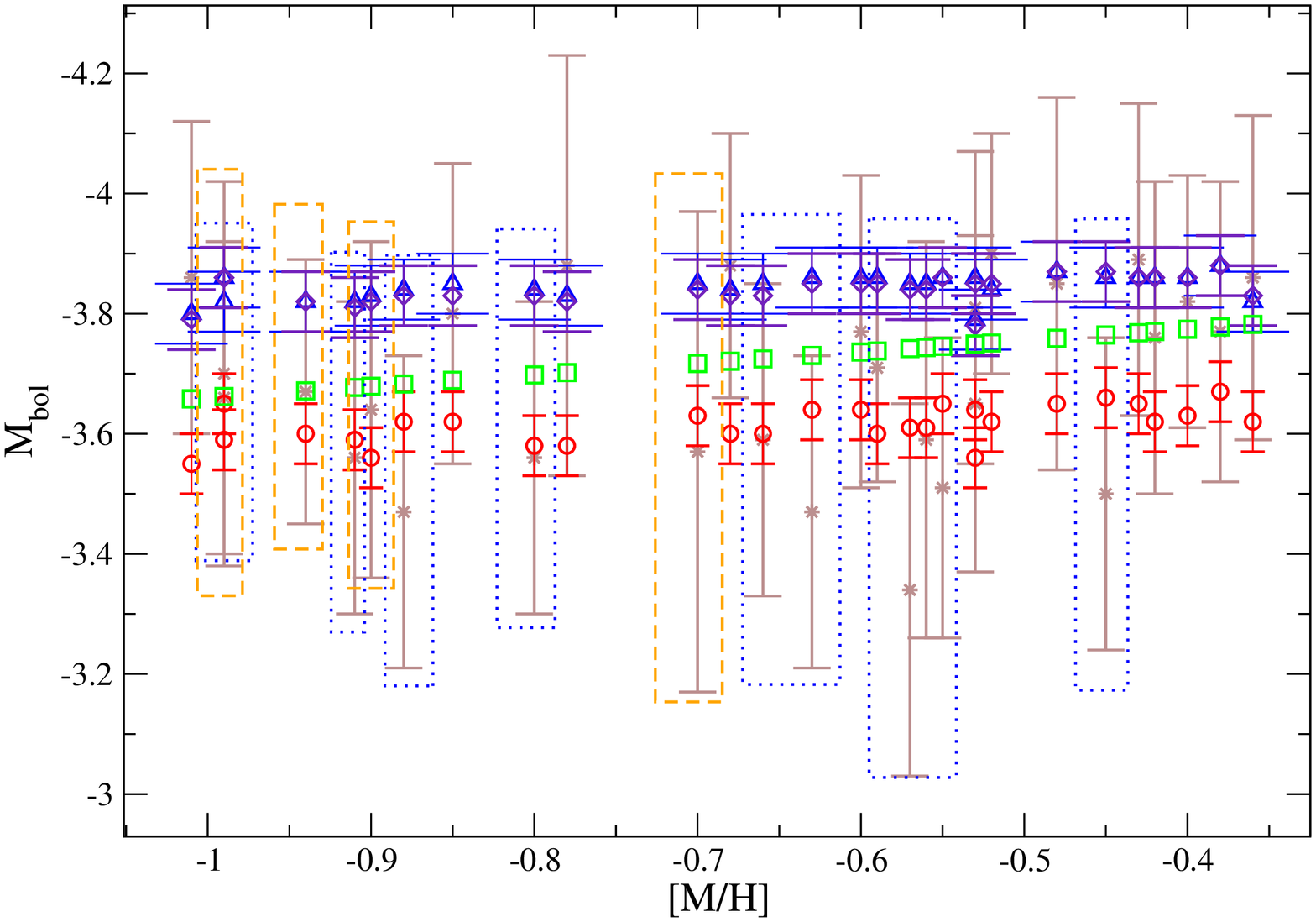}
        \end{subfigure}\vspace*{10mm}
        \begin{subfigure}{0.46\linewidth}
                \includegraphics[width=.8\linewidth, angle=-90]{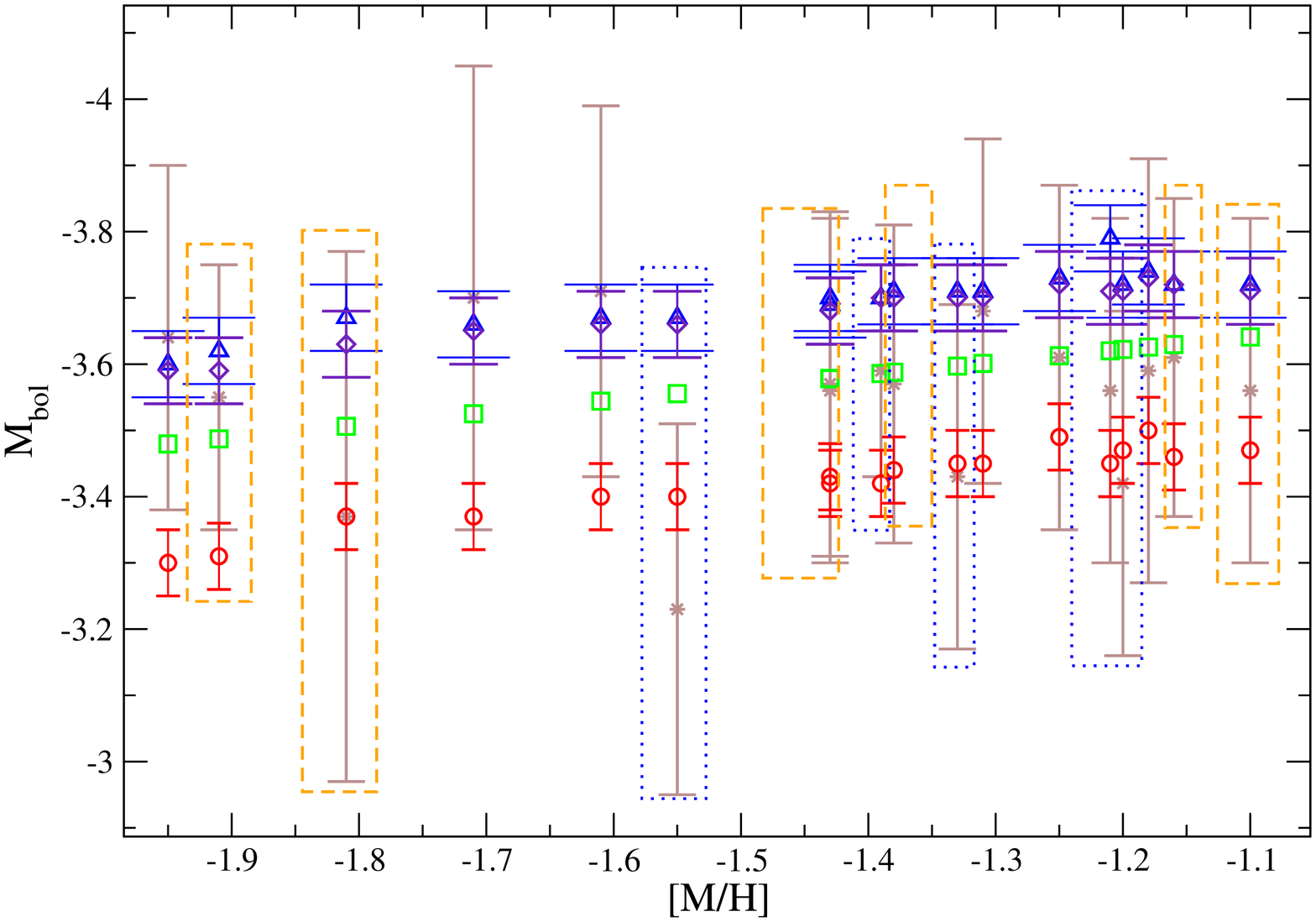}
        \end{subfigure}\hspace*{12mm}
        \begin{subfigure}{0.46\linewidth}
                \includegraphics[width=.8\linewidth, angle=-90]{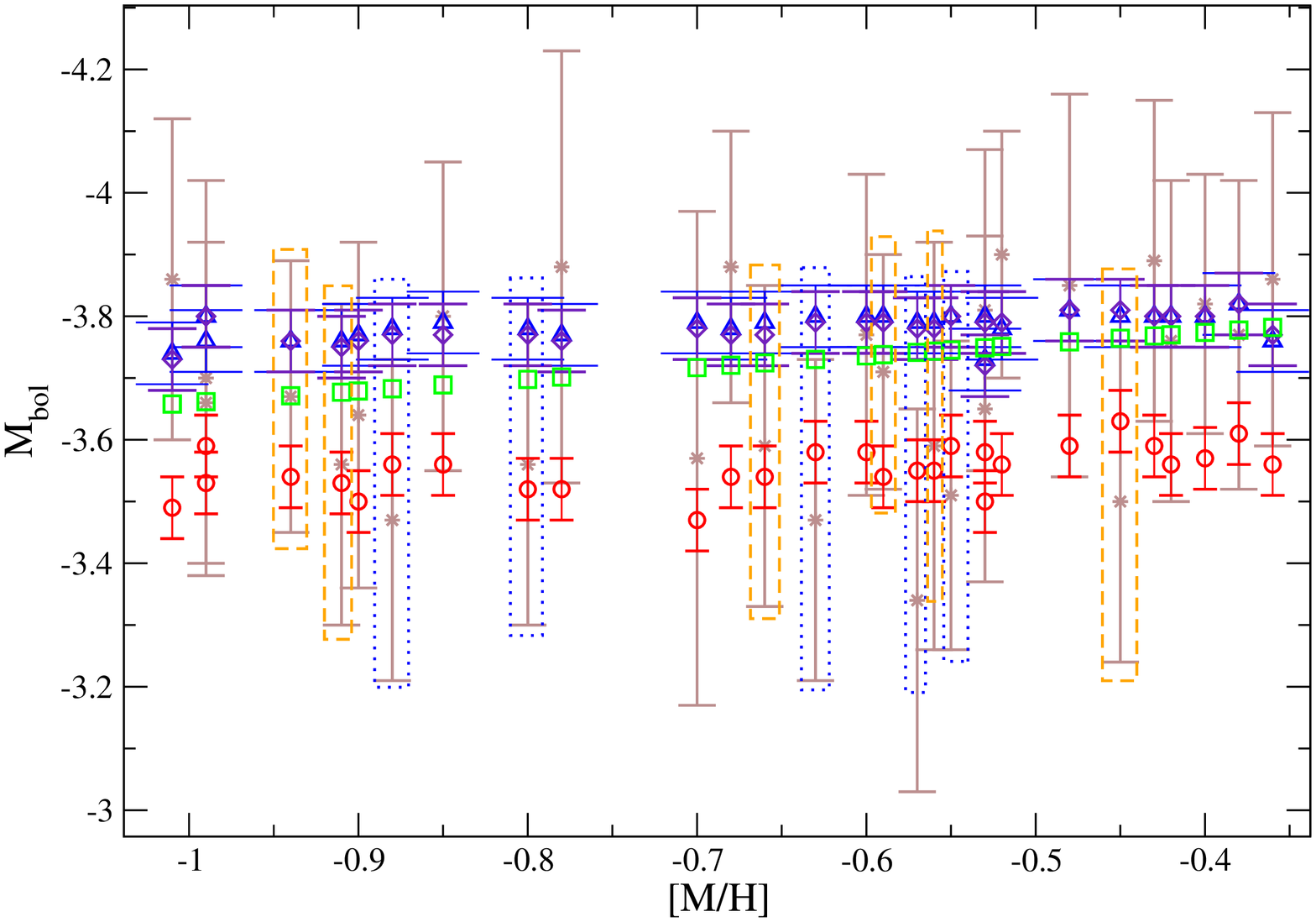}
        \end{subfigure}
         \end{figure}

The upper panels in fig. 3 compare the absolute bolometric magnitude of the globular clusters in our sample (observational calibration represented by black asterisks and empirical formula by green squares) with our stellar models (circles symbolize canonical models while triangle and diamonds correspond to models with $\mu_{12}=2.2$ and $\alpha_{26}=0.5$, respectively). The canonical models predict bolometric magnitudes closer than 0.05 from the empirical value from observations and in most cases these are less 0.1 away from the observational calibration (those clusters whose bolometric magnitude shows the best agreement are marked by asterisks). Despite being different in nature, enhanced neutrino emission and axion production raise the tip-RGB absolute magnitude by approximately the same amount (triangles almost superposed with diamonds through all the metallicity range). For 38 globular clusters, the absolute magnitude of non-standard models lies above the upper limit of the observational calibration, 30 allow to use our proposed values as constraints for the neutrino magnetic dipole moment and the axion-electron coupling constant.\\

The middle and lower panels in fig. 3 show results if the absolute bolometric magnitude of all models was shifted downwards by 0.1 and 0.16. These shifts reduce the number of clusters that allow to constraint $\mu_{12}$ and $\alpha_{26}$ to 29 and 23, respectively, by requiring higher bolometric magnitudes to overpass the error bars of observations. The confidence level is lowered down 1$\sigma$ level.


\section{Discussion and conclusions}

In this work used a sample of 50 globular clusters to get a constraint on the axion-electron coupling constant and the magnetic dipole moment of neutrinos, expanding of previous analysis \cite{ASZJ_2015B}.\\

The normality of the sample was tested by two methods:

\begin{itemize}
\item{The Anderson-Darling test, in which the hypothesis that a certain population of data, from which the sample under study is extracted, follows an specific statistical distribution gives $\mathrm{A^{2}=0.55}$. The critical value below which it can be safely assumed that the sample depicts a normal distribution is 0.75.}
\item{The histogram of the sample (left panel in fig. 3) closely resembles a normal distribution, in which the bin with the largest frequency corresponds to $\mathrm{M_{bol}^{tip}=-3.64}$. Only two globular clusters have a absolute bolometric magnitude similar to the one predicted by stellar models with non-standard energy losses.}
\end{itemize}

A further test on the influence of metallicity over the calibration for $\mathrm{M_{bol}^{tip}}$, the sample was divided into two smaller ones separated at $\mathrm{[M/H]=-0.99}$ and used to construct the histogram shown in the right panel in fig. 3. While the sub-sample with $\mathrm{[M/H]<-0.99}$ has a distribution closely resembling the complete sample (with the majority of clusters having $\mathrm{M_{bol}^{tip}=-3.59}$) the one with higher metallicity also shows a considerable number of globular clusters with $\mathrm{-3.90\leq M_{bol}^{tip}\leq -3.80}$. This points to a possible systematic brightening the overall estimation for $\mathrm{M_{bol}^{tip}}$, as these clusters also show relatively large error bars (some large enough to include the absolute bolometric magnitude for the tip-RGB predicted by nonstandard models.\\

The sample used in this work, an expansion of the one used in \cite{ASZJ_2015B}, allows to get the constraints $\mathrm{\mu_{\nu}\leq2.2\times10^{-12}\mu_{B}}$ and $\mathrm{\alpha_{26}\leq0.5\times10^{-26}}$ within a confidence level around 1.8-$\sigma$. These constraints, supported by robust statistics, hold up even if the overall absolute bolometric magnitude of the tip-RGB of stellar models is shifted downwards by 0.1 mag due to uncertainties on conventional physical ingredients. Among the several theoretical uncertainties in these constraints, the initial helium content, the nuclear reaction rates and electron conductivity and the initial mass are the most important \cite{ASZJ_2015B}. The initial amount of helium affects the bolometric luminosity of the tip-RGB: an increment by 30\% in Y$_{i}$ requires $\mathrm{\mu_{\nu}\leq2.6\times10^{-12}\mu_{B}}$ and $\mathrm{\alpha_{26}\leq0.6\times10^{-26}}$ to surpass the observational limits. The new conductive opacities and the N14+p reaction rates could induce a systematic dimming of the tip-RGB by about 0.08 and 0.12 mag. In less degree, initial mass also affects the tip luminosity (e.g. for an stellar model with M$_{i}=1.1$M$_{\odot}$ the tip is lower by about 0.04 mag). The overall uncertainty of stellar models, 0.06 over our metallicity grid, combined with the last two factors lead to a shift of about 0.16 magnitudes.\\

On the observational side, the uncertainties come from the distance modulus, reddening and the statistical uncertainty, due to the intrinsically low population of the last two magnitude bins of the RGB of most clusters. Future surveys, probably with the James Webb telescope, could greatly improve the empirical calibration for bolometric magnitude of the tip-RGB.


\section*{Acknowledgments}
We gratefully acknowledge travel support by the bilateral Conacyt-DFG
grants No. 121554 and 147751.

{}

\end{document}